\documentclass[aps,prl,twocolumn,groupedaddress,reprint]{revtex4-1}
\usepackage{graphicx}
\usepackage{dcolumn}
\usepackage{amssymb}
\usepackage{color}

\begin{document}

\newcommand{\note}[1]{\textcolor[rgb]{0.2, 0.7, 0.3}{\textsc{[#1]}}}
\newcommand{\todo}[1]{\textcolor[rgb]{0.7, 0.2, 0.2}{$=>$ #1}}
\newcommand{\theme}[1]{\textcolor[rgb]{0.2, 0.2, 0.7}{\underline{#1}}\\}
\newcommand{\chem}[1]{\ensuremath{\mathrm{#1}}}
\newcommand{\m}[1]{\mathrm{#1}}
\renewcommand{\u}[1]{\,\mathrm{#1}}
\newcommand{\etal}{\emph{et al}.\ }
\newcommand{\tocite}{\textcolor[rgb]{1.0, 0.4, 0.3}{[?]}}
\renewcommand{\vec}[1]{\mathbf{#1}}
\newcommand{\articletitle}{Sparse Cyclic Excitations Explain the Low Ionic Conductivity of Stoichiometric Li$_{7}$La$_{3}$Zr$_{2}$O$_{12}$}

%\preprint{}

\title{\articletitle}


\author{Mario Burbano$^{1,2,3}$}
\author{Dany Carlier$^{2,3}$}
\author{Florent Boucher$^4$}
\author{Benjamin J.Morgan$^5$}
\author{Mathieu Salanne$^{1,3}$}
%\thanks{}
\affiliation{$^1$Sorbonne Universit\'{e}s, UPMC Univ Paris 06, CNRS, UMR 8234, PHENIX, F-75005 Paris, France.}
\affiliation{$^2$CNRS, Universit\'{e} de Bordeaux, ICMCB, 87 avenue du Dr A. Schweitzer, 33608 Pessac Cedex, France.}
\affiliation{$^3$R\'{e}seau sur le Stockage Electrochimique de l'Energie (RS2E), FR CNRS 3459, France}
\affiliation{$^4$Institut des Mat\'{e}riaux Jean Rouxel (IMN), Universit\'{e} de Nantes, CNRS, 2 rue de la Houssini\`{e}re, BP 32229, 44322 Nantes cedex 3, France.}
\affiliation{$^5$Department of Chemistry, University of Bath, Claverton Down BA2 7AY, United Kingdom}
%\author{Dany Carlier}
%\affiliation{CNRS, Universit\'{e} de Bordeaux, ICMCB, 87 avenue du Dr A. Schweitzer, 33608 Pessac Cedex, France.}
%\affiliation{R\'{e}seau sur le Stockage Electrochimique de l'Energie (RS2E), FR CNRS 3459, 33 rue Saint Leu, 80039 Amiens Cedex, France}
%\phone{+123 (0)123 4445556}
%\fax{+123 (0)123 4445557}
%\author{Florent Boucher}
%\author{Benjamin J. Morgan}
%\author{Mathieu Salanne}
%\altaffiliation{R\'{e}seau sur le Stockage Electrochimique de l'Energie (RS2E), FR CNRS 3459, 33 rue Saint Leu, 80039 Amiens Cedex, France}
%\affiliation{Sorbonne Universit\'{e}s, UPMC Univ Paris 06, CNRS, UMR 8234, PHENIX, F-75005 Paris, France.}
%\affiliation{Maison de la Simulation, USR 3441, CEA - CNRS - INRIA - Universit\'{e} Paris-Sud - Universit\'{e} de Versailles, F-91191 Gif-sur-Yvette, France}
\email{mathieu.salanne@upmc.fr}


\date{\today}


\begin{abstract}
We have performed long time-scale molecular dynamics simulations of the cubic and tetragonal phases of the solid lithium-ion--electrolyte Li$_{7}$La$_{3}$Zr$_{2}$O$_{12}$ (LLZO), using a first-principles parameterised interatomic potential. Collective lithium transport was analysed by identifying dynamical excitations; persistent ion displacements over distances comparable to the separation between lithium sites, and string-like clusters of ions that undergo cooperative motion. We find that dynamical excitations in c-LLZO are frequent, with participating lithium numbers following an exponential distribution, mirroring the dynamics of fragile glasses. In contrast, excitations in t-LLZO are both temporally and spatially sparse, consisting preferentially of highly concerted lithium motion around closed loops. This qualitative difference is explained as a consequence of lithium ordering in t-LLZO, and provides a mechanistic basis for the much lower ionic conductivity of t-LLZO compared to c-LLZO. 
\end{abstract}


\maketitle

Conventional lithium-ion batteries rely on unstable liquid-organic polymer electrolytes, which pose practical limitations in terms of flammability, miniaturization, and safe disposal. A possible solution is to replace liquid electrolytes with inorganic ceramics that are electrochemically stable and non-flammable. The family of garnet-like oxides with general formula $\m{Li}_xM_3M^\prime_2\m{O}_{12}$, where $M$ = La and $M^\prime$ = Nb, Ta or Zr, have attracted significant attention in this regard due to their high lithium-ion conductivity, high electrochemical stability window, and chemical stability with respect to metallic lithium \cite{ThangaduraiEtAl_ChemSocRev2014, ThangaduraiEtAl_JPhysChemLett2015}. The highly stuffed garnet \chem{Li_7La_3Zr_2O_{12}} (LLZO) is the most studied member of this family, and can be considered prototypical. LLZO exhibits two phases with strikingly different ionic conductivities: a cubic phase (c-LLZO) that is adopted at high temperature ($>600\u{K}$) or stabilized by doping \cite{BernsteinEtAl_PhysRevLett2012,kuhn_li_2011}  with $\sigma\approx10^{-4}\u{S}\u{cm}^{-1}$, and a tetragonal (t-LLZO) phase with $\sigma\approx10^{-6}\u{S}\u{cm}^{-1}$ that is favoured in the pure system at ambient temperature. 

\begin{figure}[tb]
  \centering
    \includegraphics[width=0.43\textwidth]{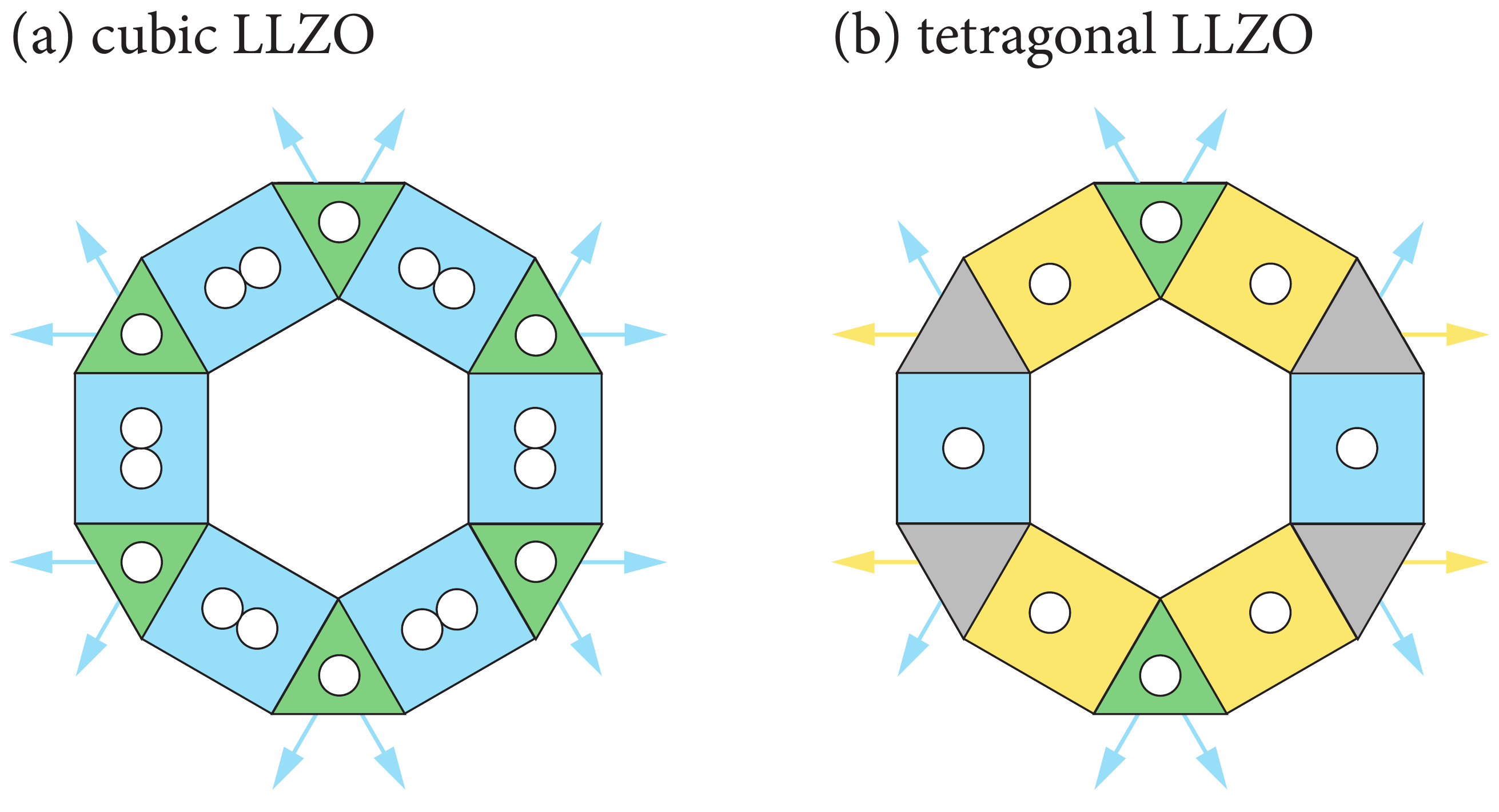}
  \caption{Schematic of the polyhedral ring topologies in (a) cubic and (b) tetragonal LLZO. Triangles represent tetrahedra, and rectangles represent octahedra. Arrows indicate neighbouring octahedra within conjoined rings. In c-LLZO lithium resides at off-center positions in the octahedra, and is disordered over all available octahedra and tetrahedra. In t-LLZO lithium is ordered over two tetrahedra and four octahedra in each ring. White circles show these lithium sites, which are partially occupied in c-LLZO.}
  \label{fig:LLZO_loops}
\end{figure}

The large difference in ionic conductivity between c-LLZO and t-LLZO is interesting from a mechanistic perspective because the pathways available for lithium transport are topologically identical in the two phases. Lithium ions move through an open three dimensional network of rings. Each ring comprises twelve alternating tetrahedral and octahedral sites (Fig.\ \ref{fig:LLZO_loops}) \cite{AwakaEtAl_ChemLett2011}, and the tetrahedral sites act as nodes connecting adjacent rings. In stoichiometric LLZO each ring accommodates on average eight lithium ions, which preferentially occupy all six octahedra and two of the tetrahedra. In the cubic phase the tetrahedral sites are equivalent and the lithium ions are disordered. In the tetragonal phase the tetrahedra are inequivalent due to the reduced crystal symmetry, and lithium occupies tetrahedral pairs aligned along the $\left[001\right]$ direction forming an ordered sublattice \cite{BernsteinEtAl_PhysRevLett2012, kuhn_li_2011}. This lithium ordering is correlated with the sharp decrease in ionic conductivity relative to the cubic phase. A mechanistic explanation of the relationship between lattice symmetry, lithium ordering, and ionic transport is, however, lacking.

Previous theoretical studies identified that lithium diffusion in c-LLZO and t-LLZO phases occurs via correlated ionic motion \cite{AdamsAndRao_JMaterChem2012, JalemEtAl_ChemMater2013, MeierEtAl_JPhysChemC2014}. The relationship between microscopic ion-correlations and macroscopic lithium transport, characterised by diffusion coefficients and ionic conductivities, is however unresolved. Meier \etal used \emph{ab initio} molecular dynamics (MD) and metadynamics simulations to study cubic and tetragonal LLZO \cite{MeierEtAl_JPhysChemC2014}. These authors observed diffusion in c-LLZO effected by asynchronous correlated hopping of lithium ions, while in t-LLZO a synchronous concerted process involving multiple lithium ions was identified. In contrast, an earlier \emph{ab initio} MD study of c-LLZO by Jalem \etal attributed diffusion in the cubic phase to synchronous lithium hopping \cite{JalemEtAl_ChemMater2013}. The apparently inconsistent descriptions from these studies illustates the challenge of adequately sampling ionic transport in materials with complex correlated diffusion mechanisms. Direct first principles simulations are computationally costly, and these two studies considered relatively short simulation trajectories ($10$--$30\u{ps}$) that captured only a few diffusion events \footnote{in the case of the Meier \etal study, in the low ionic conductivity tetragonal phase only a single diffusion event was observed, even with the enhanced configuration space sampling provided by metadynamics.}. This limited sampling means it is not clear whether individual diffusion processes represent long-time ensemble diffusion behaviour. If we wish to connect microscopic lithium dynamics with macroscopic transport behaviour, much longer simulations and a statistical treatment of the resulting trajectories are necessary.
 
Here we report long time-scale classical MD simulations of cubic and tetragonal LLZO, with particular emphasis given to analysing their room temperature lithium transport. These simulations use an interatomic potential (IP) model for lithium garnets derived from state-of-the-art hybrid Density Functional Theory (h-DFT) calculations. This allows much longer simulation times than direct \emph{ab initio} calculations at moderate computational expense, while avoiding empirical parameterisation \cite{RotenbergEtAl_PRL2010}. We have characterised lithium transport in both phases using two techniques typically used to study glass-forming liquids \cite{keys_excitations_2011, donati_stringlike_1998}, and not previously applied to solid electrolytes. This analysis reveals qualitatively different transport statistics for c-LLZO versus t-LLZO. We identify displacive \emph{excitations}, defined as persistent ion displacements over distances comparable to the separation between lithium sites. Excitation events are frequent in c-LLZO, but highly \emph{sparse} in t-LLZO. A statistical comparison of lithium diffusion behaviour is made by identifying string-like clusters of ions that undergo cooperative motions \cite{donati_stringlike_1998}. In c-LLZO the probability distribution of string lengths is exponential, mirroring the behaviour of fragile glassy liquids \cite{donati_stringlike_1998,vogel_temperature_2004}. In contrast, the string length distribution in t-LLZO is \emph{discontinuous}, with distinct preferred string lengths. The most probable string lengths correspond to cyclic cooperative motions of lithium ions around closed loops. These cyclic processes give zero net displacement of charge, and therefore do not contribute to ionic conductivity. The low conductivity of t-LLZO is therefore attributed to persistent diffusive processes being sparse in location and time, and dominated by zero-charge displacement closed-loop cyclic processes.
  
Classical MD allow simulations across the length- and timescales necessary to describe collective ionic motion at a fraction of the computational cost of \textit{ab initio} methods. This efficiency often comes at the expense of accuracy. One solution is to use physically motivated IPs that reproduce the electron density response of individual ions to changes in their coordination environment \cite{madden_covalent_1996}. Parameters for the resulting models can be derived from electronic structure (i.e.\ DFT) calculations, to produce IPs that are accurate and transferable \cite{burbano_ceria_2014}. Such IPs are capable of describing changes in composition and local structure, such as those which occur during ionic transport, or across families of similar materials. Parameters for a DIPole Polarizable Ion Model (DIPPIM) \cite{madden_covalent_1996}, used throughout this work, were obtained by calculating sets of forces, dipole moments, and stresses using h-DFT across a sample of different stoichiometries and atomic geometries, and then minimising the least squares errors between the h-DFT and DIPPIM data. Although computationally expensive, h-DFT functionals give improved structural parameters, such as lattice constants, compared to standard DFT \cite{da_silva_hybrid_2007, burbano_ceria_2014}. A full description of the parameterization procedure and validation of the resulting IP against experimental structural data is given in the Supplementary Information.      

We performed molecular dynamics simulations of the lithium garnet \chem{Li_7La_3Zr_2O_{12}}, using a $2\!\times\!2\!\times\!2$ supercell containing 1536 atoms, with a timestep of $1\u{fs}$. The system was equilibrated for temperatures ranging between $300\u{K}$ and $1000\u{K}$ in the isothermal-isobaric (NPT) ensemble. The supercells were initially equilibrated at a temperature of $280\u{K}$ for 10 ps; the temperature was then scaled up to $1000\u{K}$ at a rate of $1\u{K}\u{ps}^{-1}$. Production runs were performed in the canonical (NVT) ensemble using the equilibrium values from the NPT simulations and were up to $87.4\u{ns}$ long, in the case of t-LLZO at $300\u{K}$.

\begin{figure}[htb]
\centering
\includegraphics[width=0.45\textwidth]{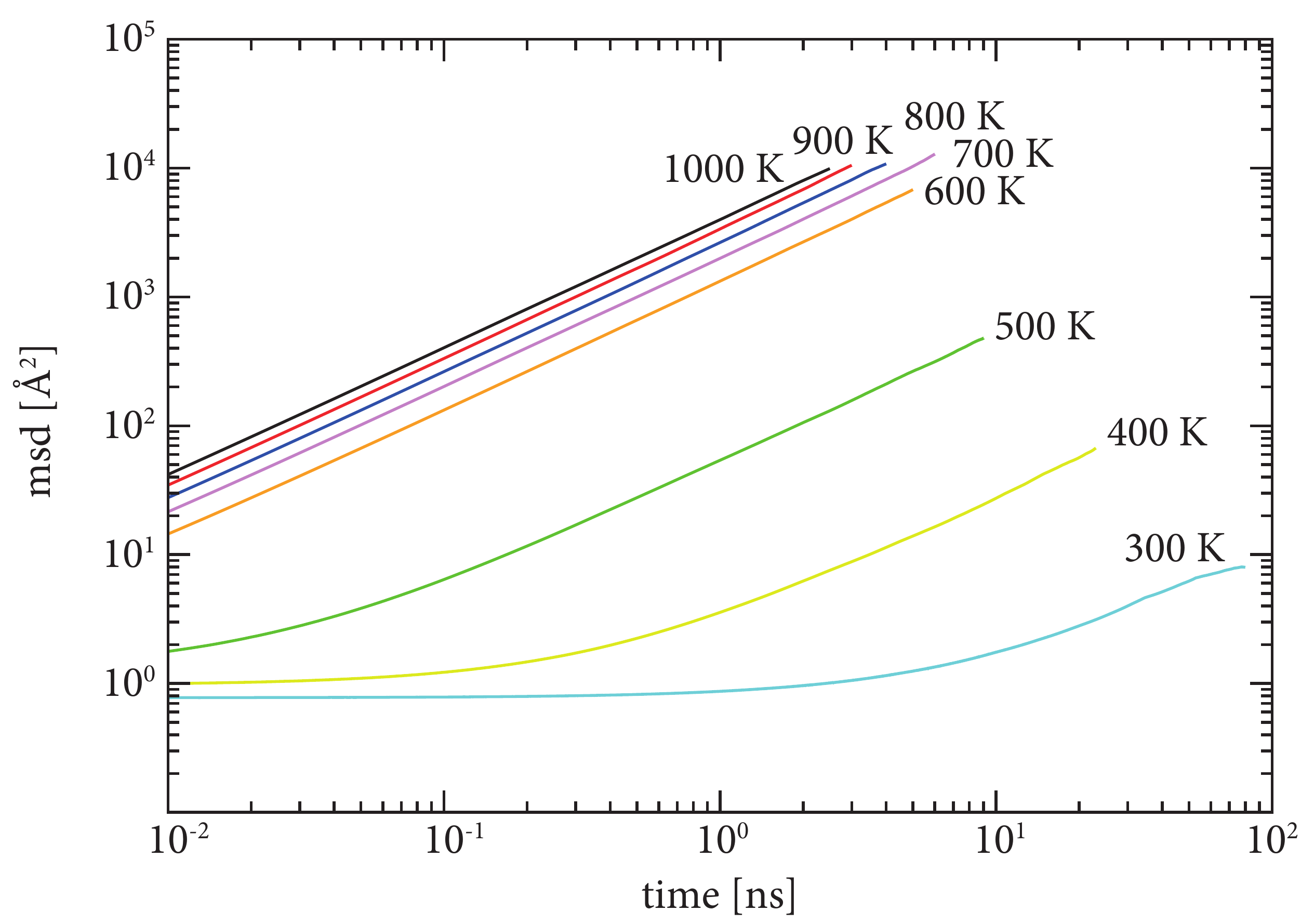}
  \caption{Calculated mean-squared displacements for LLZO at $300$--$1000\u{K}$. On a log--log plot, the diffusive regime corresponds to the sloping linear portion.}%, with the diffusion coefficient proportial to the point on the $y$ axis from extrapolating this region. }
  \label{fig:msds}
\end{figure}

Fig.\ \ref{fig:msds} shows lithium mean-squared displacements calculated at temperatures from $300$ to $1000\u{K}$ \cite{MorganAndMadden_JPCM2012}. At high temperatures, where the c-LLZO phase is adopted, lithium diffusion is fast and the diffusive regime is sampled even by short simulation trajectories. In contrast, below $600\u{K}$, where the tetragonal phase is favoured, the msd is orders of magnitude smaller. To sample the diffusive regime well, long-time simulations become necessary, which would be prohibitively costly for \emph{ab initio} methods.

To characterise lithium transport in materials with concerted ionic diffusion, one cannot simply examine hopping frequencies of individual ions (as for simple interstitial or vacancy diffusion), but must instead consider the collective motions of groups of atoms and the contributions made by these to ensemble transport coefficients \cite{MorganAndMadden_PhysRevLett2014}. Providing that a system is in a particle hopping regime (i.e.\ is not ``superionic'' \cite{Catlow_SSI1983a}) local diffusion events are well separated in time. Net contributions to ensemble transport correspond to non-trivial particle displacements and exclude vibrational motion or very short lived configurational changes. We identify transitions between relatively long-lived configurations, or ``excitations'', using the methodology of Keys \etal for supercooled glass forming liquids \cite{keys_excitations_2011}. In this case, mobile ions involved in excitations are selected using the following functional of particle trajectories:
\begin{equation}
h_{i}\left(t,t_{a};a\right) =  \prod_{t^{\prime} = t_{a}/2-\Delta t}^{t_{a}/2} \theta\left(\left|\mathbf{{r}}_{i}\left(t+t^{\prime}\right) - \mathbf{{r}}_{i}\left(t-t^{\prime}\right)   \right| - a \right),
\label{eq:prod}
\end{equation}
where $a$ is a displacement cutoff, $\Delta t$ is the typical time for a mobile particle to move to a distinct position, here $5\u{ps} $, and $t_a>\Delta t$ is a sufficiently long time to allow a complete transition between microstates, here set to $30\u{ps}$. $\theta$ is the Heaviside step function, $\theta(x) = 1$ or $0$ for $x\ge0$ or $<0$, respectively. The product is evaluated for each lithium ion at every frame of the simulation trajectory. Summing over all particles each frame then gives the number of particles involved in each excitation event. For our analysis of both LLZO phases, $a=3\u{bohr}$, which is approximately the Li--Li intersite distance.

\begin{figure}[htb]
\centering
\includegraphics[width=0.47\textwidth]{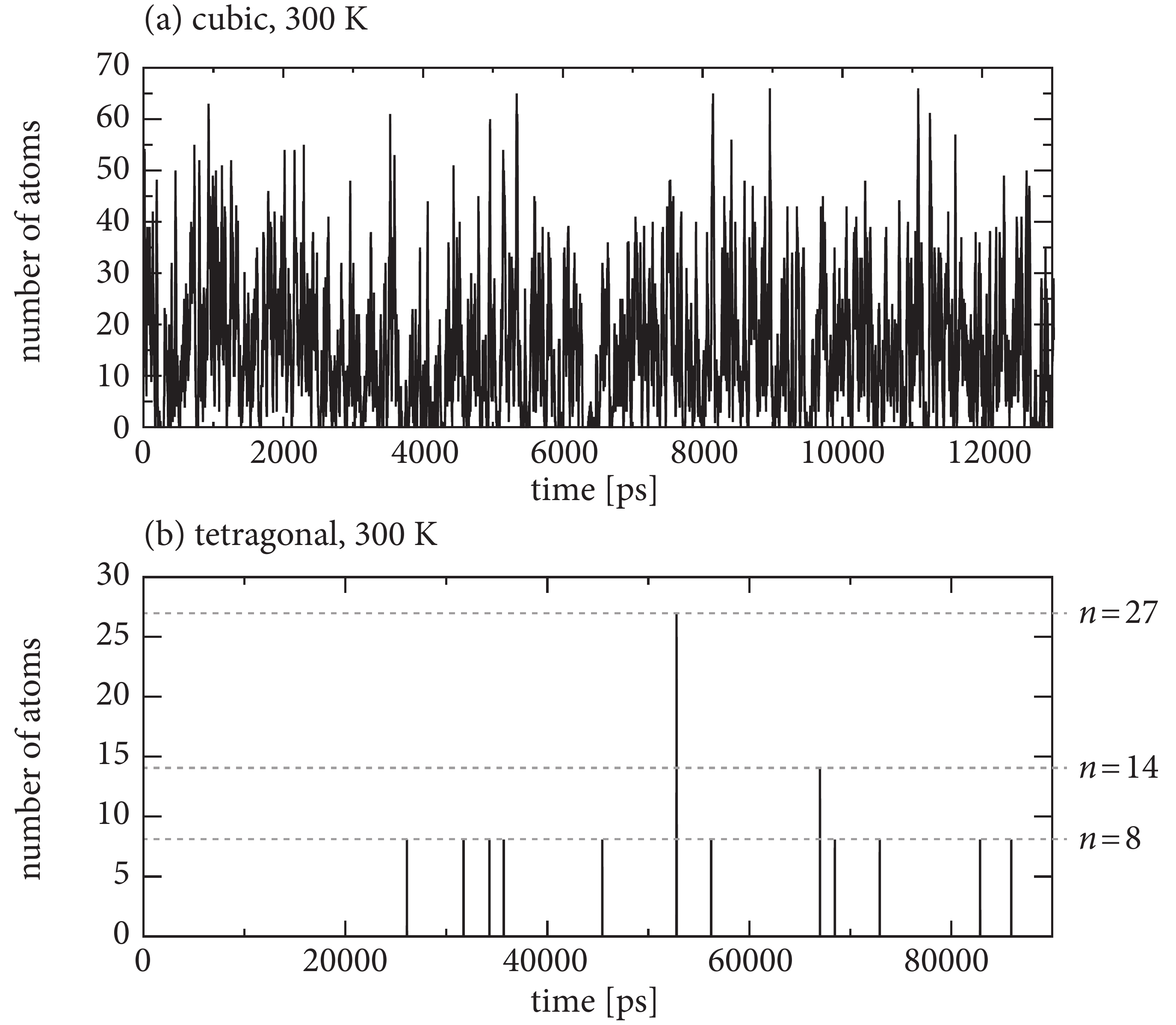}
  \caption{Excitation statistics at $300\u{K}$ for cubic (upper panel) and tetragonal (lower panel) LLZO.}
  \label{fig:sumh}
\end{figure}

To allow a direct comparison between phases, a c-LLZO cell was also simulated at $300\u{K}$ by enforcing a cubic lattice shape. Excitation statistics for c-LLZO and t-LLZO at $300\u{K}$ are shown in Fig.\ \ref{fig:sumh}. Each peak corresponds to an excitation event, where the peak height is the number of participating lithium ions. For c-LLZO, even at $300\u{K}$ there are many excitations, with well-distributed numbers of contributing atoms. The t-LLZO excitation statistics, however, are strikingly different. Even at the long simulation timescale considered ($87.4\u{ns}$), few excitations are observed. Collections of ions move in close succession, and many nanoseconds pass between events with no persistent diffusive motion. The t-LLZO data also show a strong preference for specific excitiation sizes. Ten of the twelve observed excititions involve exactly eight lithium ions, suggesting a specific eight-atom process dominates lithium dynamics.% in the tetragonal phase.

As a supporting analysis of the lithium diffusion statistics in c-LLZO and t-LLZO, we identify ``strings'' of lithium involved in cooperative diffusion, using a procedure proposed by Donati \etal for studying dynamics in supercooled liquids \cite{donati_stringlike_1998}. The original procedure generates strings by connecting pairs of mobile particles $i$ and $j$ if 
\begin{equation}
  \min\left( \left| \vec{r}_i(t) - \vec{r}_j(0) \right|,\left| \vec{r}_j(t) - \vec{r}_i(0) \right|\right) < r_{\textrm{min}}.
\end{equation}
In a solid-lithium electrolyte, such as LLZO, lithium ions move between well defined coordination tetrahedra and octahedra. Applying this concept of lithium hopping between lattice sites allows a modified connection criterion of
\begin{equation}
    S_i(t+\Delta t) = S_j(t),
\end{equation}
where $S_i(t)$ is the site occupied by ion $i$ at time $t$. The set of oxygen coordinates at each time step define instantaneous polyhedra geometries, which allow us to assign every lithium to a specific lattice site. To reduce noise from short-lived vibrational motions (which do not contribute to diffusion) we filter changes in site occupation for each lithium that are reversed at the next time step. 

\begin{figure}
\includegraphics[width=0.45\textwidth]{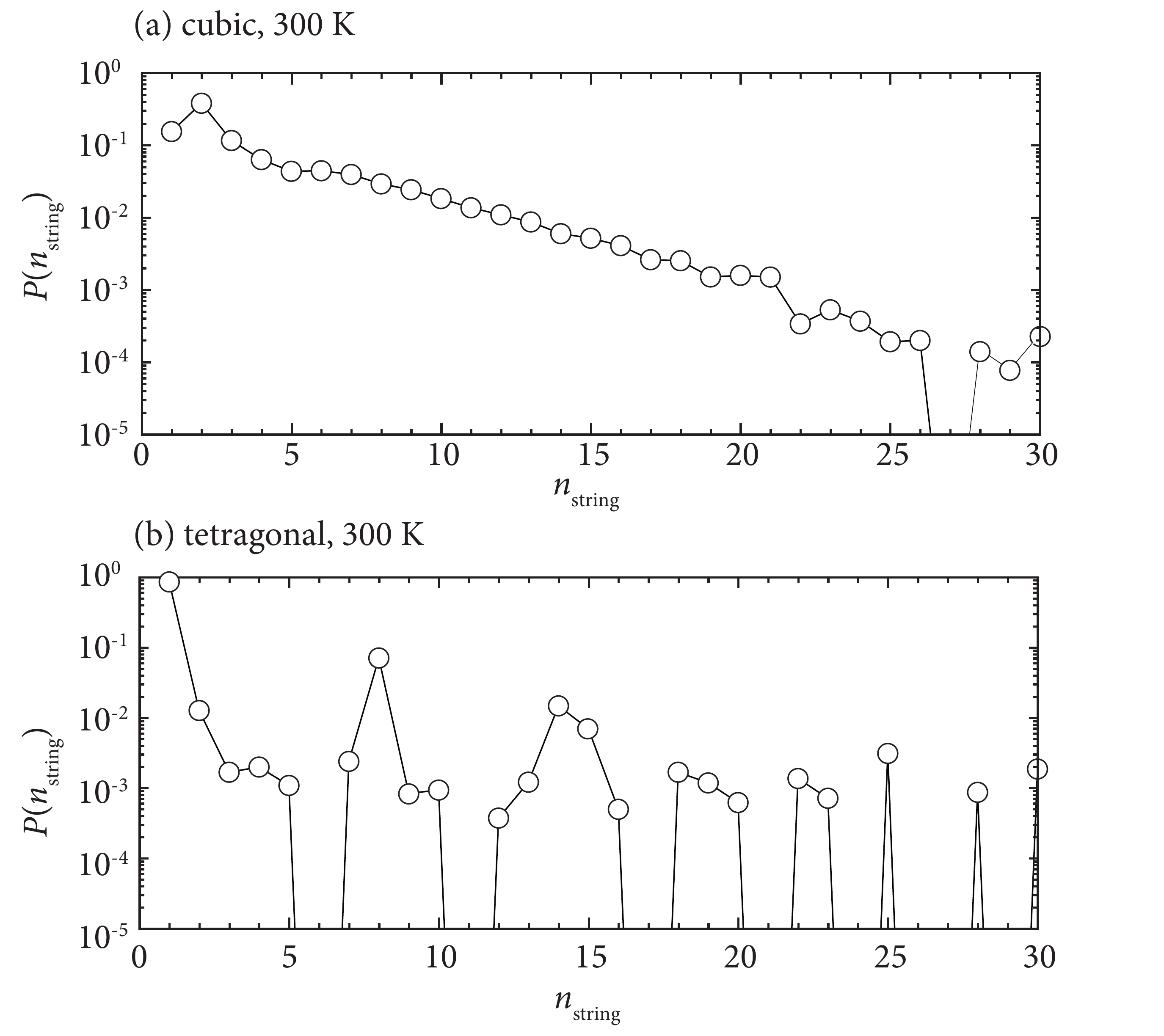}
  \caption{Probability distribution, P($n$) of string lengths $n$ for c-LLZO (upper) and t-LLZO (lower). $\Delta t=30\u{ps}$, for consitency with the excitation analysis.}
  \label{fig:strings}
\end{figure}

Fig.\ \ref{fig:strings} shows the probability distributions, $P(n_\m{string})$, that a diffusing Li ion contributing to a string of length $n_{\m string}$. Echoing the differences in excitation statistics, the string statisics for the two phases are qualitatively different. For c-LLZO strings of all lengths are found, and $P(n_\m{string})$ is a monotonically decreasing exponential distribution, as seen for supercooled glassy liquids \cite{donati_stringlike_1998,vogel_temperature_2004}. For t-LLZO, however, strings of length $n_\m{string}>1$ are, in general, much less likely, except for specific values of $n_\m{string}$. The most probable connected string length is $n=8$; nearly two orders of magnitude more frequent than $n=\left\{7,9\right\}$, and an order of magnitude more frequent even than $n=2$.

\begin{figure}[htb]
\centering
\includegraphics[width=0.35\textwidth]{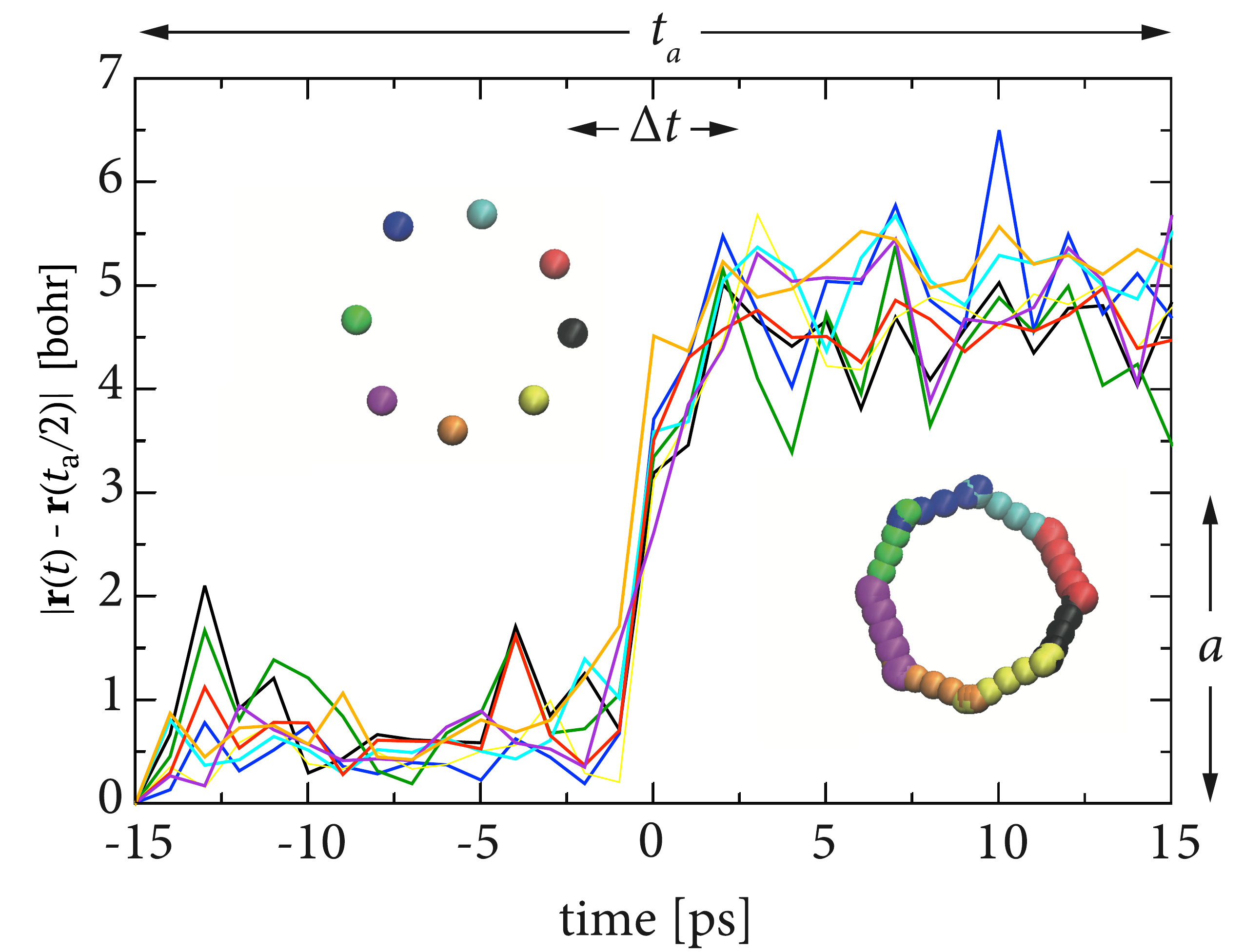}
  \caption{An example 8-ion cyclic excitation event in t-LLZO. The parameters used in the excitation functional, Eqn.\ \ref{eq:prod}, are also shown.}
  \label{fig:loop}
\end{figure}

Our analysis reveals a preference in t-LLZO for diffusion events involving eight lithium ions, that does not exist in c-LLZO. Directly examining the trajectories of ions in one of these eight-lithium diffusion events we find cooperative \emph{cyclic} processes, with eight lithium ions moving around a twelve-site ring. One such cyclic excitation is illustrated in Fig.\ \ref{fig:loop}, which shows the initial positions of the eight contributing lithium ions alongside their trajectories and individual displacements. Longer processes that involve more lithium ions are also found. The t-LLZO excitation statistics reveal a single excitation with $n=14$, which matches the second peak in $P(n_\m{string})=14$. These 14 lithium processes are also closed loops, that involve coherent lithium motion around two rings. Another excitation involving 27 ions is also observed, which extends across the simulation cell boundary before meeting with its periodic image (Fig. S2). 

The strong preference for eight-lithium cyclic diffusion in t-LLZO can be understood as a consequence of the tetragonal lattice symmetry and associated lithium ordering \cite{BernsteinEtAl_PhysRevLett2012}. The stoichiometry of LLZO gives exactly eight lithium ions per ring. Each ring has six octahedral and six tetrahedral sites. In t-LLZO the lithium preferentially occupies all six octahedral sites and two of the tetrahedral sites on opposite sides of the ring \cite{AwakaEtAl_ChemLett2011}. The tetrahedral sites are nodes connecting rings and the orientation of occupied tetrahedral sites in one ring is therefore correlated with its neighbours, producing a fully ordered phase (ignoring thermal disorder) \cite{BernsteinEtAl_PhysRevLett2012}. In an ordered phase an arbitrary set of mobile ions moving between sites may be classified according to whether the initial and final configurations are (locally) equivalent by symmetry. We distinguish \emph{complete} processes, where the start and end points are equivalent, from \emph{partial} diffusion processes. In t-LLZO, a complete diffusion process conserves the arrangement of occupied tetrahedral sites. From this perspective, the eight-lithium cyclic excitations can be explained as the shortest possible complete diffusion process. Longer complete processes are also possible---for example, concerted diffusion of 14 lithium ions around a closed loop of two rings---but less likely.

The much lower ionic conductivity of t-LLZO compared to c-LLZO can therefore be attributed to two factors. First diffusive excitations in t-LLZO are highly \emph{sparse}, and occur much less frequently than in c-LLZO. Secondly, excitations in t-LLZO are predominantly isolated cyclic processes around closed loops. These have zero net charge displacement, and therefore do not contribute to ionic conductivity. If only closed cyclic diffusion processes occurred in t-LLZO the ionic conductivity would be zero. We can speculate that in a macroscopic system occasional excitations will meet and coalesce with a second excitation, then a third, and on, thus allowing ionic conduction. The 27 ion excitation we observe shows an analogue of this behaviour. This excitation crosses a cell boundary and connects with itself because of the simulation periodic boundary conditions.

In conclusion, we have performed long-time MD simulations of cubic and tetragonal LLZO using a first principles based IP. Our long simulation times allow thorough sampling of the lithium diffusion behaviour, even in the poorly conducting t-LLZO phase. By applying statistical analyses, previously used to study transport in glassy materials, we find lithium diffusion in c-LLZO and t-LLZO is qualitatively different. In c-LLZO persistent dynamical excitations are frequent and occur with exponentially distributed numbers of participating lithium ions, showing behaviour characteristic of supercooled glasses \cite{donati_stringlike_1998,vogel_temperature_2004}. In t-LLZO persistent excitations are sparse, rare events, and are dominated by concerted ionic motion around the rings that make up the lithium polyhedral network. The preference for closed loop diffusion is due to the strong lithium ordering in the tetragonal phase. The low ionic conductivity of t-LLZO compared to c-LLZO is therefore explained as a consequence of the dominant diffusive excitations in t-LLZO being sparse cyclic processes.

More generally, these results illustrate the importance of thorough statistical sampling, via long simulation trajectories, when modelling solid electrolytes that exhibit concerted ionic motion. We have shown the utility of first-principles--parameterised, transferable IPs in accurately modelling electrolyte materials over long simulation times, and how analysis techniques commonly used for studying glassy liquids can resolve statistically different diffusion behaviour in solid electrolytes.

\begin{acknowledgments}
This work was supported by the French National Research Agency (Labex STORE-EX: grant ANR-10-LABX-0076). We thank the University of Bordeaux for the use of the cluster Avakas.  We are grateful for the computing resources on OCCIGEN (CINES) obtained through the project x2015097321.  B.\ J.\ M.\ acknowledges support from the Royal Society (UF130329).
\end{acknowledgments}

\end{document}